%% file: author.tex
\documentclass{svproc}
\usepackage{url}
\usepackage{amsmath} 
\usepackage{graphicx} 

\begin{document}
\mainmatter              
\title{OptWedge: Cognitive Optimized Guidance toward Off-screen POIs}
\titlerunning{OptWedge}  
%
\author{Shoki Miyagawa}
\authorrunning{Shoki Miyagawa} 
%
\tocauthor{Shoki Miyagawa}
\institute{Information Technology R\&D Center, MITSUBISHI Electric Corporation, Kamakura 247-8501, Japan,\\
\email{Miyagawa.Shoki@ds.MitsubishiElectric.co.jp}
}

\newcommand{\argmin}{\mathop{\rm arg~min}\limits}

\maketitle              

\begin{abstract}
Guiding off-screen points of interest (POIs) is a practical way of providing additional information to users of small-screen devices, such as smart devices and head-mounted displays.
Popular previous methods involve displaying a primitive figure referred to as Wedge on the screen for users to estimate off-screen POI on the invisible vertex.
Because they utilize a cognitive process referred to as amodal completion, where users can imagine the entire figure even when a part of it is occluded, localization accuracy is influenced by bias and individual differences.
To improve the accuracy, we propose to optimize the figure using a cognitive cost that considers the influence.
We also design two types of optimization with different parameters: unbiased OptWedge (UOW) and biased OptWedge (BOW). 
Experimental results indicate that OptWedge achieves more accurate guidance for a close distance compared to heuristics approach.
\keywords{Point of Interest, Guidance, Cognitive, Bias, Optimization}
\end{abstract}
%

\input{figure}
\input{table}
\input{formula}

\input{1-introduction}
\input{2-relatedwork}

\input{3-method}
\input{4-evaluation}
\input{5-discussion}
\input{6-conclusion}

\bibliographystyle{./styles/bibtex/spmpsci}
\bibliography{reference_full} 

\end{document}

%% file: figure.tex

\newcommand{\figparam}
{
\begin{figure}[t]
    \centering
    \includegraphics[width=0.8\textwidth]{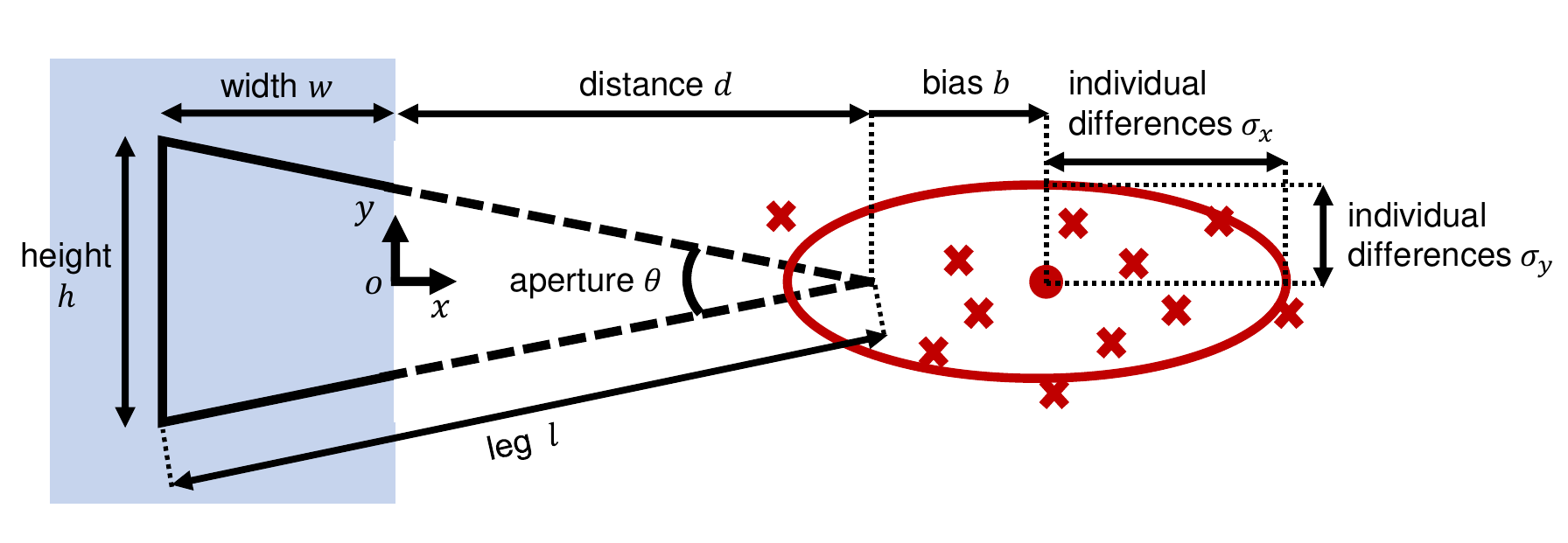}
    \caption{Parameters for determining the shape of Wedge and normal distribution $P$. The mean of the normal distribution (red dots) and range of unbiased standard deviation (red ellipses) are obtained from the human-estimated positions (indicated by a cross).}
    \label{fig:param}
\end{figure}
}

\newcommand{\figopt}
{
\begin{figure}[t]
    \centering
    \includegraphics[width=\textwidth]{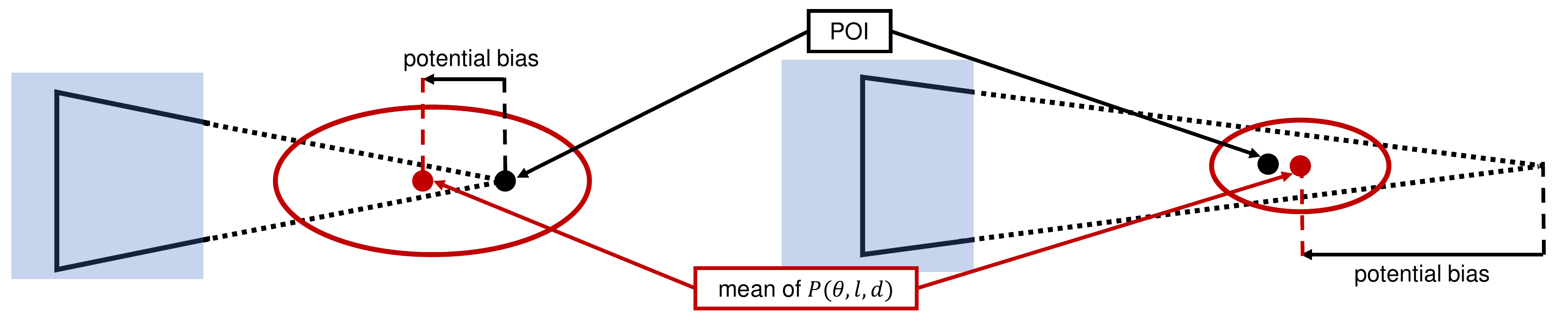}
    \caption{Comparison of shapes after optimization (lef: UOW, right: BOW).}
    \label{fig:opt}
\end{figure}
}

\newcommand{\figvr}
{
\begin{figure}[t]
    \centering
    \includegraphics[width=0.5\textwidth]{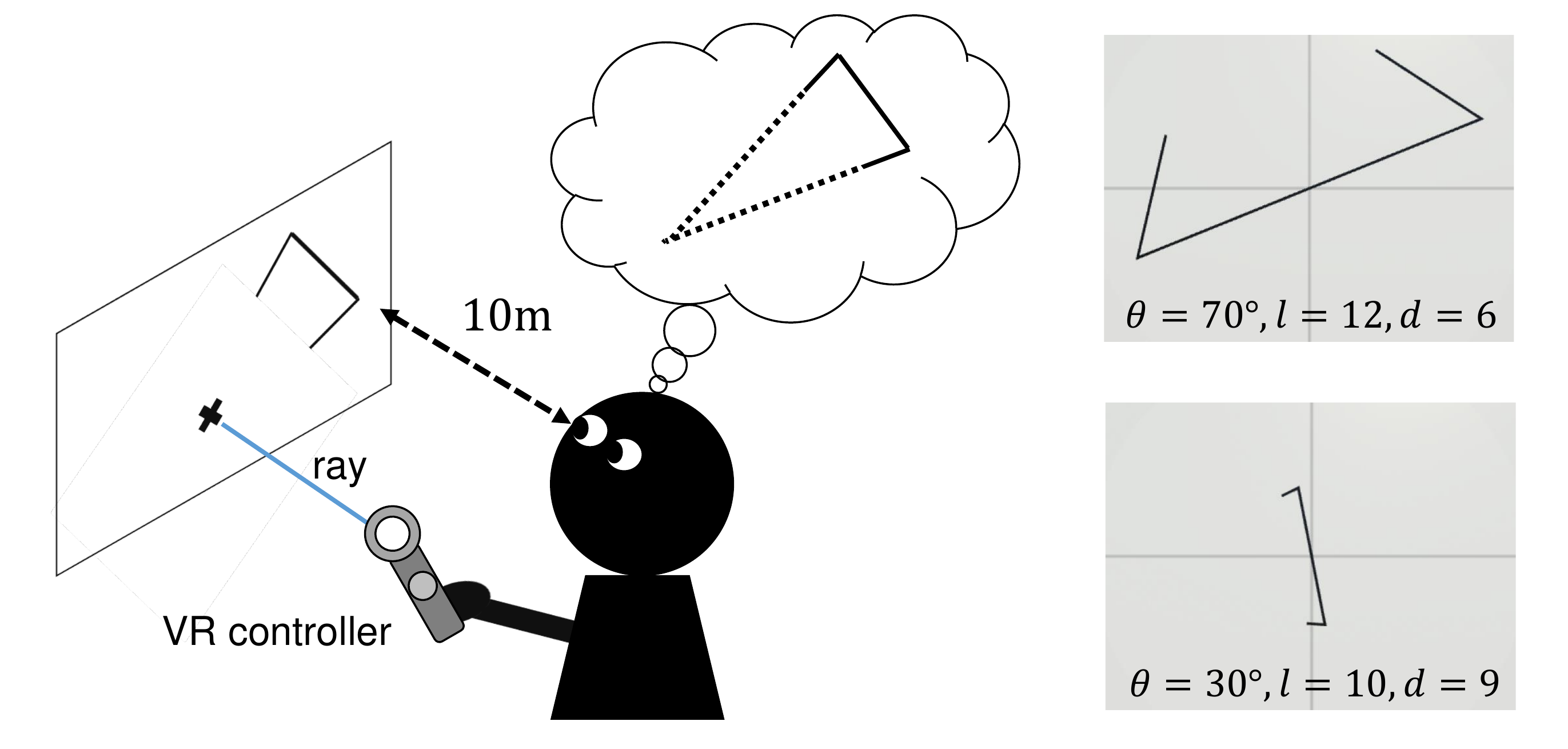}
    \caption{VR environment used in the experiment (left) and examples of parameter combinations (right).}
    \label{fig:vr}
\end{figure}
}

\newcommand{\figraw}
{
\begin{figure*}[t]
    \centering
    \includegraphics[width=\textwidth]{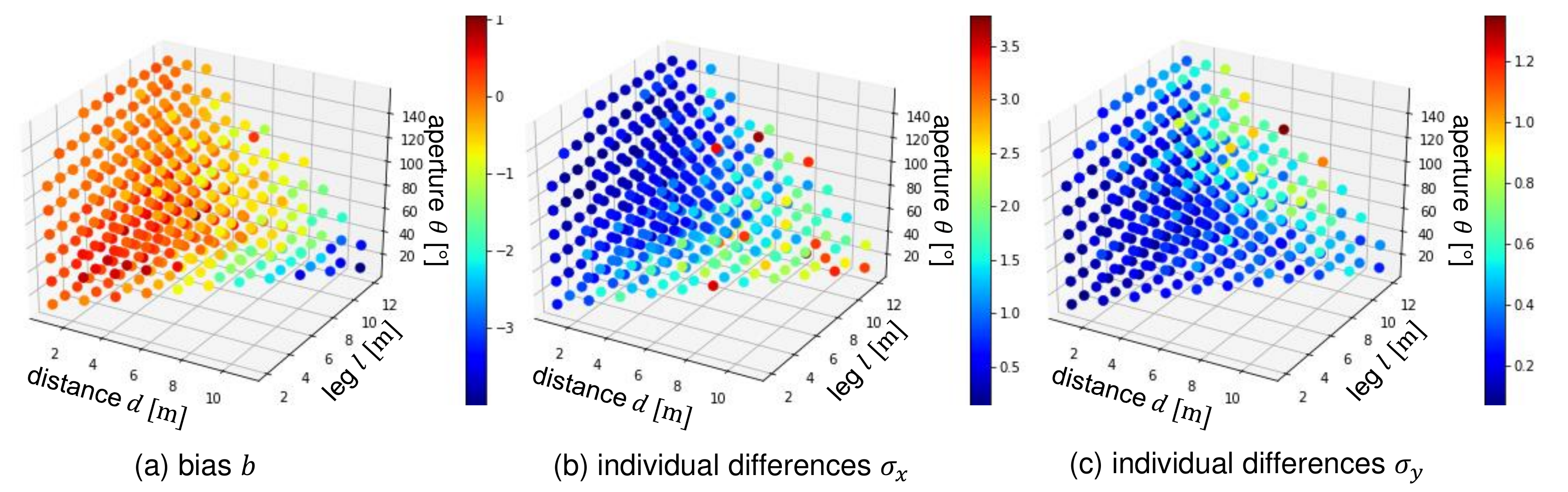}
    \caption{Relationship between parameters for Wedge shape $(\theta,l,d)$ and normal distribution $P$ $(b,\sigma_x,\sigma_y)$.}
    \label{fig:raw}
\end{figure*}
}

\newcommand{\figoffline}
{
\begin{figure}[t]
    \centering
    \includegraphics[width=0.7\textwidth]{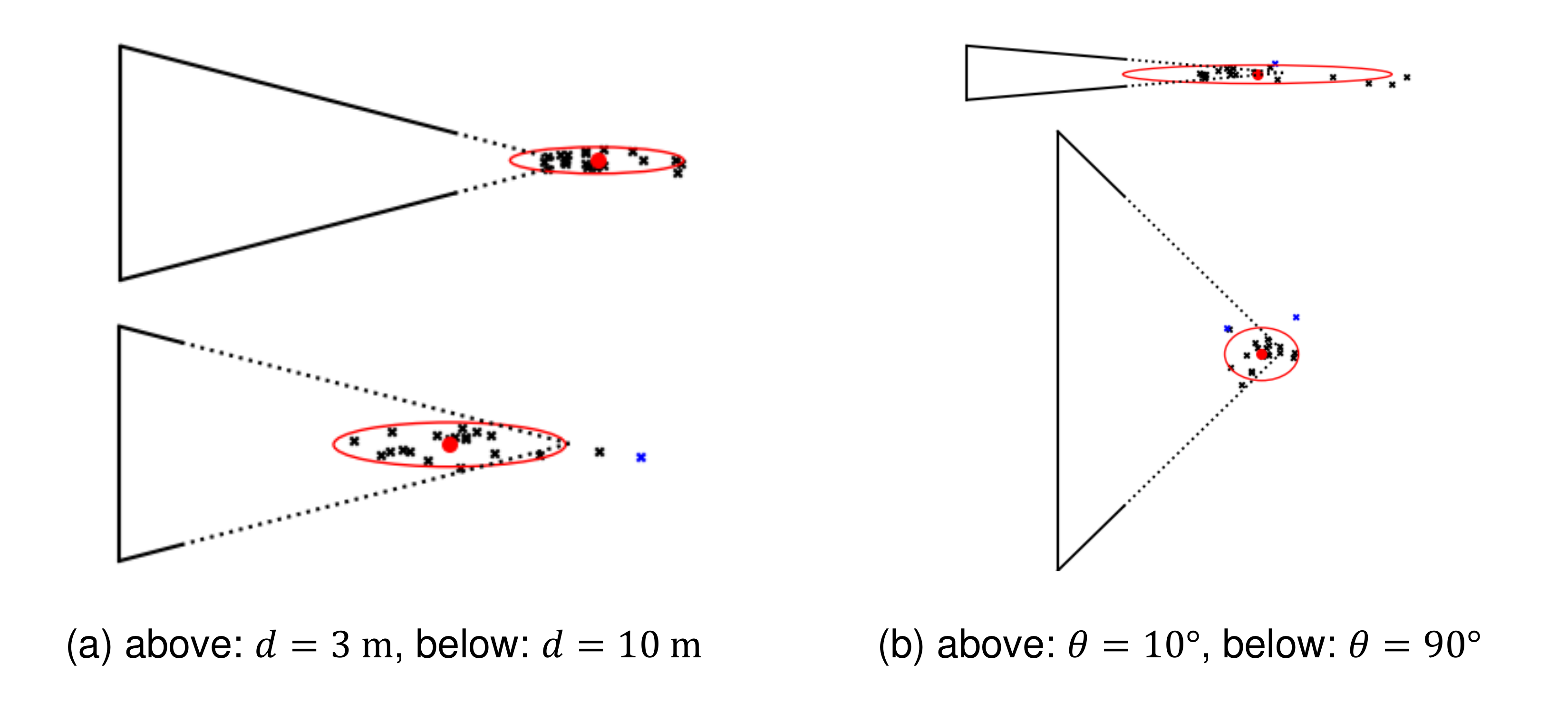}
    \caption{Comparison of distributions with (a) different $d$ (fixed for $\theta=30^\circ, l=12~\mathrm{m}$) and (b) different $\theta$ (fixed for $l=12~\mathrm{m}, d=6~\mathrm{m}$).}
    \label{fig:trend}
\end{figure}
}

\newcommand{\figcomparison}
{
\begin{figure}[t]
    \centering
    \includegraphics[width=0.5\textwidth]{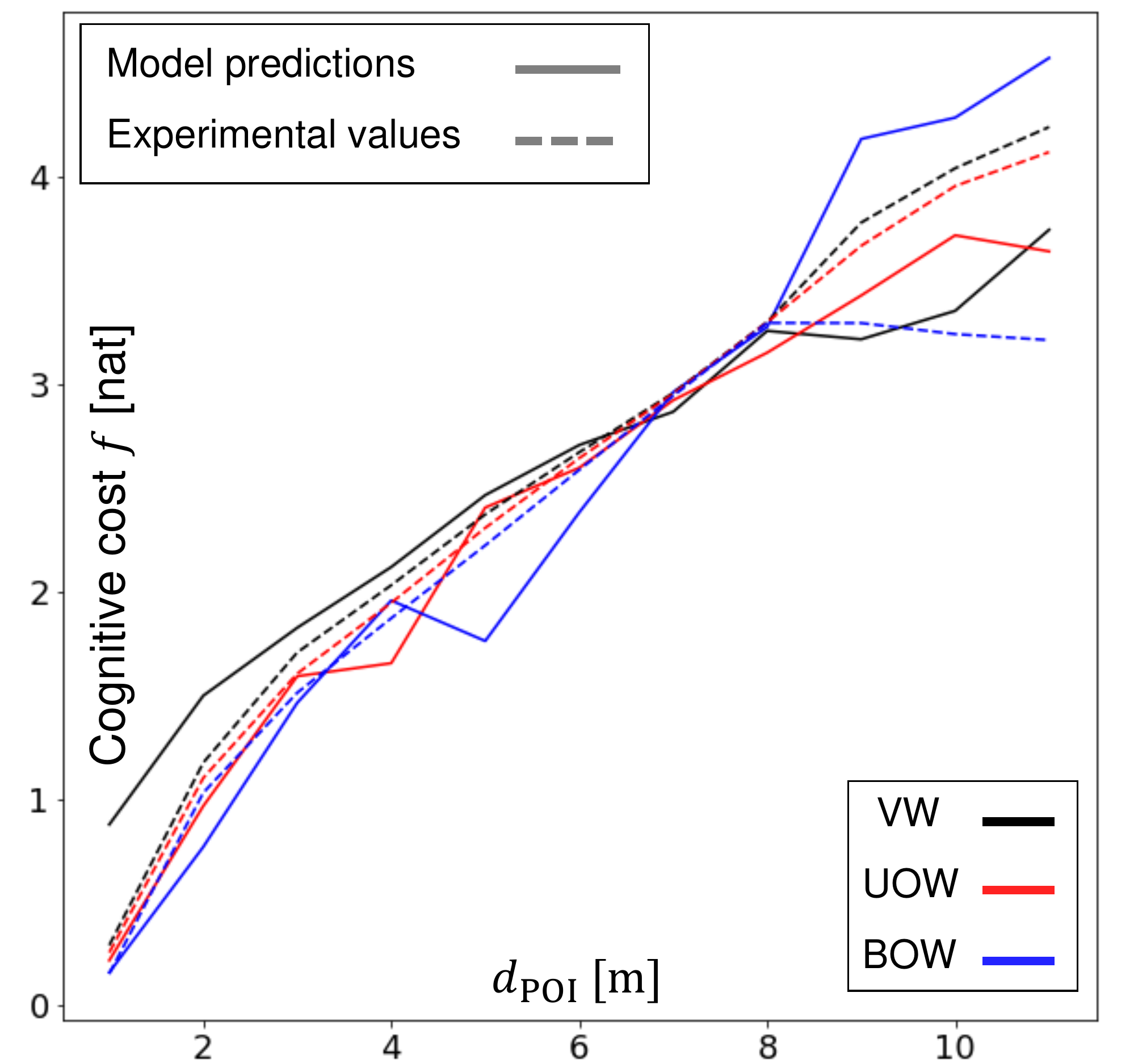}
    \caption{Relationship between distance $d_\mathrm{POI}$ to the target position and cognitive cost $f$.}
    \label{fig:comparison}
\end{figure}
}

\newcommand{\figtest}
{
\begin{figure}[t]
    \centering 
    \includegraphics[width=0.77\textwidth]{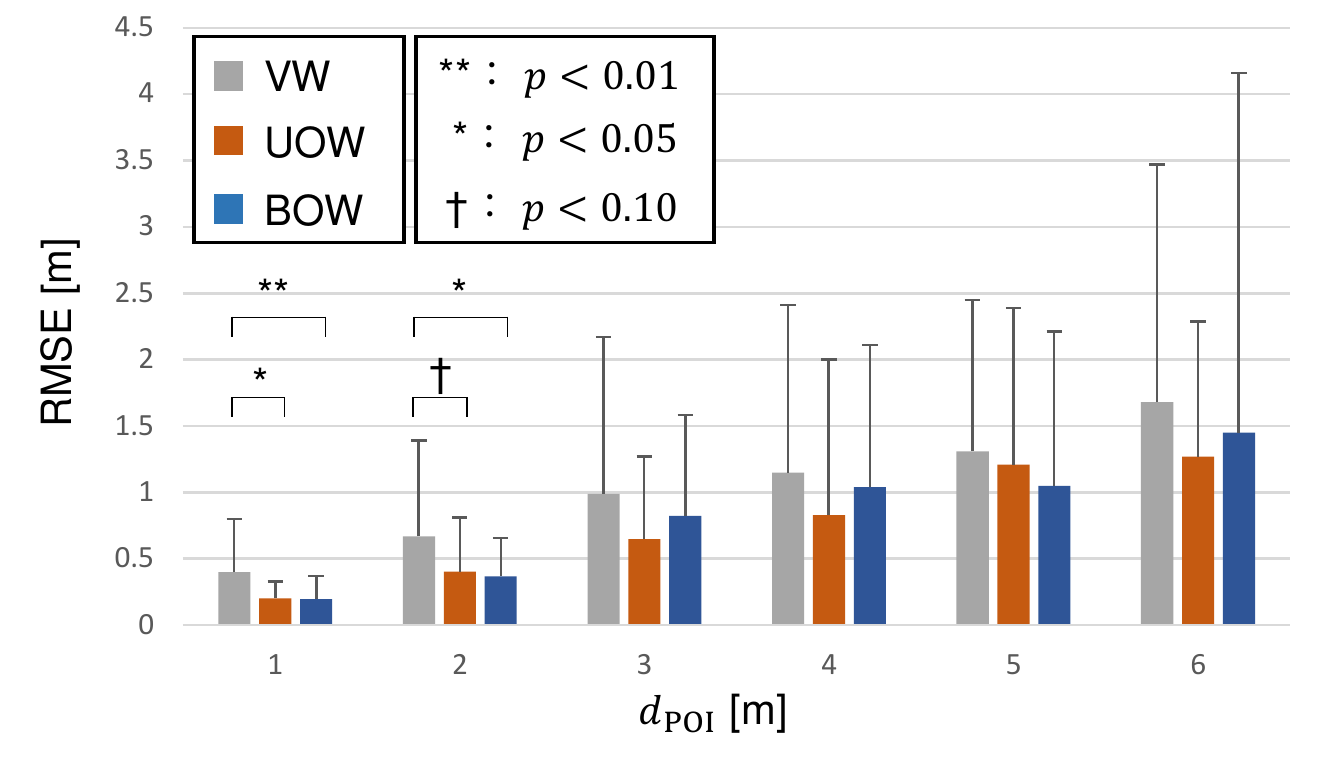}
    \caption{Results of Wilcoxon's signed-rank test (sample size $n=44$).}
    \label{fig:test}
\end{figure}
}

\newcommand{\figresults}
{
\begin{figure}[t]
    \begin{tabular}{c}
        \begin{minipage}{0.4\hsize}
            \centering
                \includegraphics[width=\textwidth]{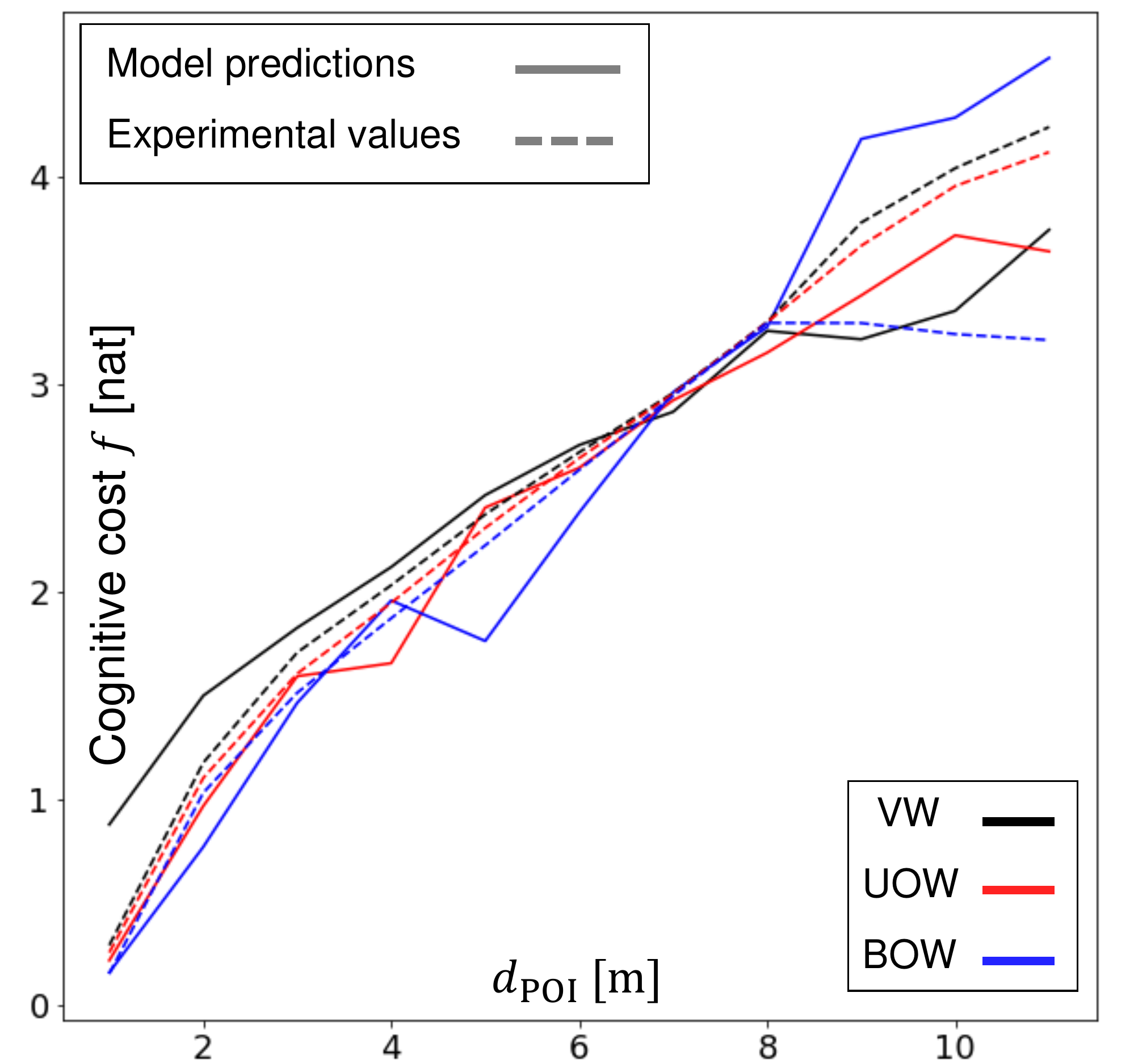}
                \caption{Cognitive cost for each Wedge.}
                \label{fig:comparison}
        \end{minipage}
        \begin{minipage}{0.6\hsize}
            \centering
                \includegraphics[width=\textwidth]{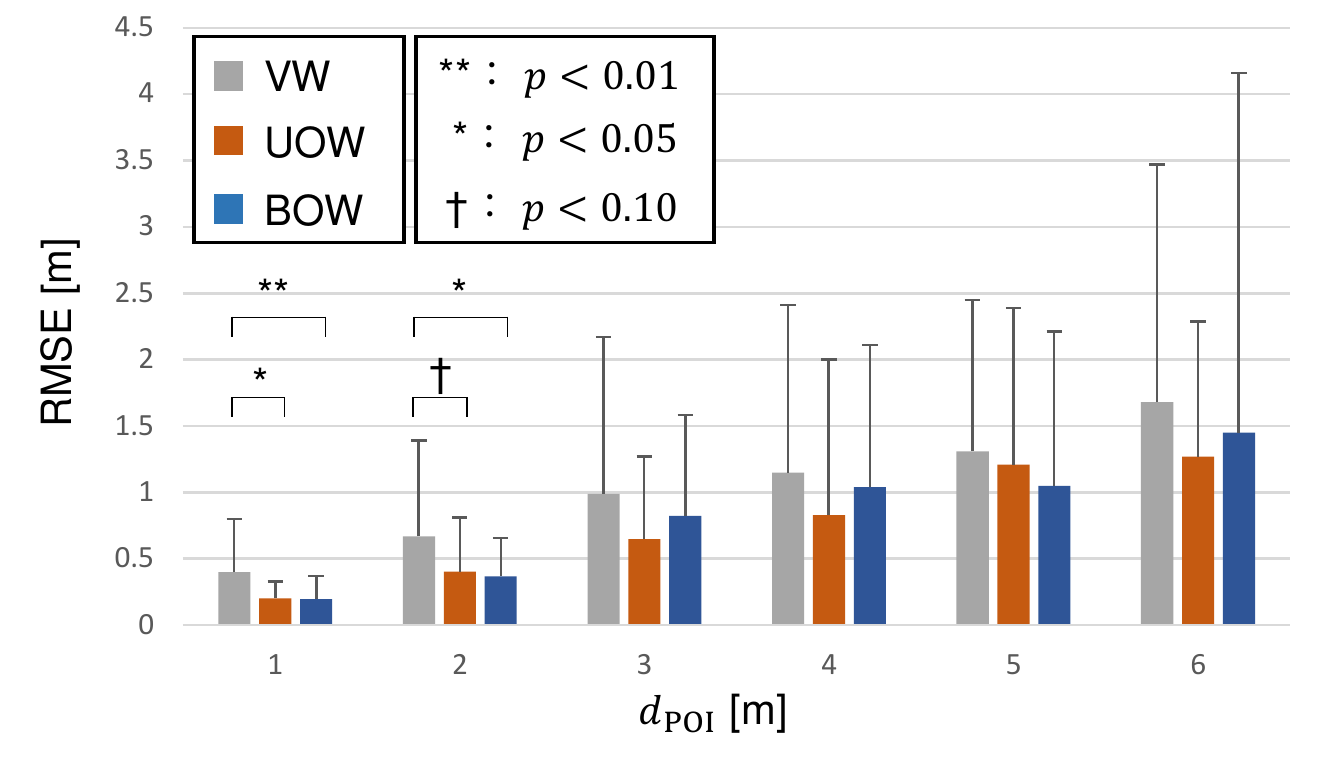}
                \caption{Wilcoxon's signed-rank test ($n=44$).}
                \label{fig:test}
        \end{minipage}
    \end{tabular}
\end{figure}
}

\newcommand{\figoptwedge}
{
\begin{figure*}[t]
    \centering
    \includegraphics[width=\textwidth]{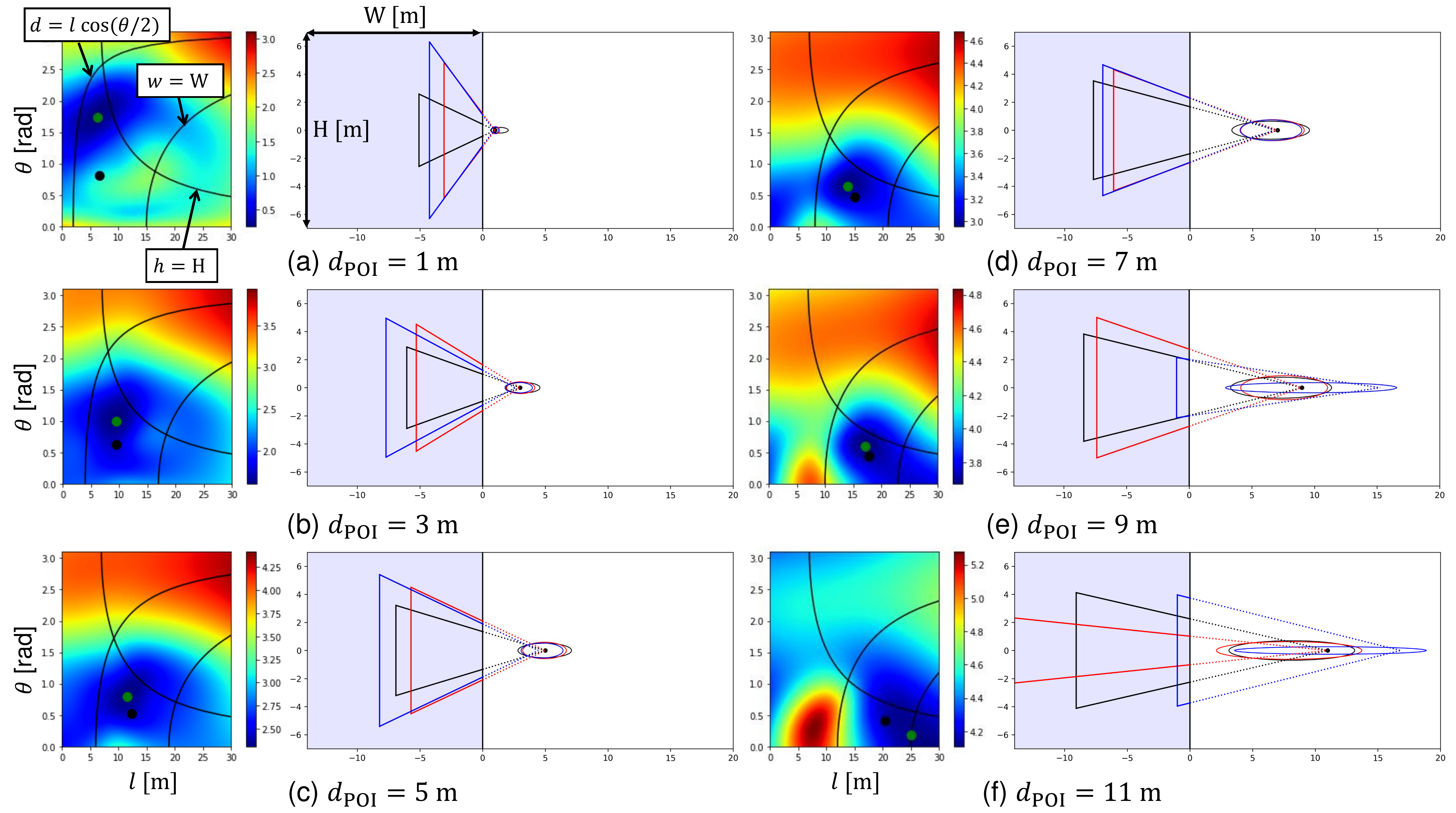}
    \caption{Comparison of the cognitive costs defined on the parameter space (left) and shape (right) corresponding to each Wedge parameter. The colors of the points and shapes in the parameter space represent each Wedge (VW: black, UOW: red, BOW: blue). The shaded blue area represents the drawable area restricted by the constraints.}
    \label{fig:optwedge}
\end{figure*}
}

%% file: table.tex
\newcommand{\tabreg}
{
\begin{table}[t]
\begin{tabular}{cc}
    \begin{minipage}[t]{0.5\hsize}
    \centering
    \caption{Comparison of \textbf{adjusted $R^2$} for polynomial regression with different orders: linear (LR), quadratic (QR), and cubic (CR). A higher value is better.}
    \label{tab:r2}
    \begin{tabular}{lcccc}
        \hline
            & LR~\cite{Gustafson08-2} & LR & QR & CR        \\ \hline
        $b$ & 0.625 & 0.643 & \textbf{0.710} & 0.676        \\
        $\sigma_x$ & 0.135 & \textbf{0.702} & 0.673 & 0.400 \\
        $\sigma_y$ & 0.080 & 0.646 & \textbf{0.764} & 0.495 \\ \hline
    \end{tabular}
    \end{minipage} 
    &
    \begin{minipage}[t]{0.5\hsize}
    \centering
    \caption{Comparison of \textbf{MSE} $[\mathrm{m}^2]$ for polynomial regression (PR) and Gaussian process regression (GR). A lower value is better.}
    \label{tab:mse}
    \begin{tabular}{lcc}
        \hline
                    & PR (best)    & GR              \\  \hline
        $b$         & 25.7  & \textbf{5.18}     \\
        $\sigma_x$  & 8.03  & \textbf{4.96}     \\
        $\sigma_y$  & 0.703 & \textbf{0.629}    \\  \hline
    \end{tabular}
    \end{minipage}
\end{tabular}
\end{table}
}

%% file: formula.tex
\newcommand{\formulaoriginal}
{
    \begin{gather*}
        l = d_\mathrm{POI} + \log\frac{d_\mathrm{POI}+20}{12}\times 10, \ \ 
        \theta(l) = (5 + d_\mathrm{POI}\times 0.3)/l
    \end{gather*}
}

\newcommand{\formulamodel}
{
    \begin{gather*}
        b = f(\theta, l, d), \ 
        \sigma_x = g(\theta, l, d), \ 
        \sigma_y = h(\theta, l, d) 
    \end{gather*}
}

\newcommand{\formulap}
{
    \begin{gather*}
        P(\theta,l,d) = \mathcal{N}\left(
            \left[
            \begin{array}{c}
                d+b(\theta, l, d) \\
                0 
            \end{array}
            \right]
            ,
            \left[
            \begin{matrix}
                \sigma_x^2(\theta, l, d) & 0 \\
                0 & \sigma_y^2(\theta, l, d)
            \end{matrix}
            \right]
        \right)
    \end{gather*}
}

\newcommand{\formulaq}
{
    \begin{gather*}
        \mathrm{Q} = \mathcal{N}\left(
            \left[
            \begin{array}{c}
                d_\mathrm{POI} \\
                0 
            \end{array}
            \right]
            ,
            \left[
            \begin{matrix} 
                \mathrm{\epsilon_x^2} & 0 \\
                0 & \mathrm{\epsilon_y^2}
            \end{matrix}
            \right]
        \right) \\
    \end{gather*}
}

\newcommand{\formulakld}
{
    \begin{equation*}
        f(\theta, l, d) = D_{KL}(\mathrm{Q} || P) = \int_x\int_y\mathrm{Q}\log\frac{\mathrm{Q}}{P(\theta, l, d)}dxdy
    \end{equation*}
}

\newcommand{\formulaopt}
{
\begin{gather*}
    \begin{array}{lc}
    (\mathrm{UOW}) & 
        (\hat{\theta}, \hat{l}) = \argmin_{\theta, l}\left(
            f(\theta, l, d_\mathrm{POI}) + \sum_i^{N} g_{i}(\theta, l, d_\mathrm{POI})
        \right) \\
    (\mathrm{BOW}) & 
        (\hat{\theta}, \hat{l}, \hat{d}) = \argmin_{\theta, l, d}\left(
            f(\theta, l, d) + \sum_i^{N} g_{i}(\theta, l, d)
        \right)
    \end{array}
\end{gather*}
}

\newcommand{\formulacons}
{
    \begin{equation*}
        g(x)= \sum_i p_iF_i(x) + \sum_j q_jG_j(x)\quad (p_i,q_j\leq 0)
    \end{equation*}
}

\newcommand{\formulaconsoption}
{
    \begin{gather*}
        g_1(\theta, l, d) = w - \mathrm{W} = l\cos(\theta/2)-d - \mathrm{W} \leq 0 \\
        g_2(\theta, l) = h - \mathrm{H} = 2 \times l\sin(\theta/2) - \mathrm{H} \leq 0
    \end{gather*}
}

\newcommand{\formularange}
{
    \begin{gather*}
        \theta\ [^\circ] \in\left\{10,30,\ldots,130, 150\right\} \\
        l\ [\mathrm{m}]\in\{2,3,\ldots,11,12\} \\
        d\ [\mathrm{m}]\in\{1,2,\ldots,10,11\}
    \end{gather*}
}

%% file: 1-introduction.tex
\section{Introduction}
Guiding points of interest (POIs) involves providing information to users on where they should pay attention.
This type of guidance generally helps users to conduct specific tasks, such as navigation on a map~\cite{Miau16}, and watching a 360-degree video~\cite{Lin17}.
However, POIs are not always visible because of the limitations of the screen size.
When the screen is as small as a smartwatch, most POIs are located off-screen.
Alternatively, if we regard the screen as the human eye, the POI may be outside the screen even when the actual screen size is room-scaled~\cite{Petford19}.
Regardless of the screen size, some POIs may be located off-screen and not visible because users usually do not know the locations of the POIs in many applications.
Although zooming and panning interactions make POIs visible, this operation is time consuming and troublesome as the number of POI increases.
Therefore, many researchers have designed methods that visualize off-screen POIs and guide them effectively.

There are several performance indicators in guidance.
For example, a search time required for guidance is vital for emergency guidance.
Some papers have modeled~\cite{Ens11} and improved~\cite{Gruenefeld19-1}\cite{Petford19} this search time.
User preference is also necessary in terms of the system usability.
These high performances assume that guidance is so accurate that the error from the POI to the guided position is sufficiently small.
However, we found few discussions to improve specific guidance to achieve accurate localization, although many researchers have focused on emphasizing either multiple POIs comparison or task-planning.
Therefore, we focused on particular guidance and tried to improve the localization accuracy in this study.

Some guidance makes users estimate the location of the POI, resulting in a high cognitive load.
For example, popular previous methods use geometrical figures such as a circle (named \textit{Halo}~\cite{Baudisch03}) and an isosceles triangle (named \textit{Wedge}~\cite{Gustafson08-1}), and display a part of it near the screen edge.
These methods utilize cognitive processing referred to as amodal completion~\cite{Kellman91}, where users can imagine the entire figure even partially occluded.
In the case of Halo, users can imagine an invisible center of the circle based on its visible arc.
In Wedge, users can estimate an invisible vertex by extending two sides that are partially visible and intersected virtually.
There happen localization errors because of two cognitive factors: bias and individual differences.
Our motivation is to reduce these negative impacts and facilitate more accurate estimation.
However, to the best of our knowledge, no previous study has considered the influence of cognitive processing, including amodal completion.
Because they usually determined parameters of Halo and Wedge heuristically, it is unclear whether the parameters are reasonable or not.

Therefore, we propose a cognitive guidance that explicitly controls cognitive influences to improve existing methods.
Specifically, we set an optimization problem where figure-related parameters are optimized by minimizing the cognitive cost.
This method allows us to check the validity of the parameters as well as introduce prior knowledge on the application.
For example, the figure should not be too large because it is minor content on the screen.
Conventional heuristic approaches cannot explicitly handle such constraints; however, our methods can introduce these constraints into the optimization problem.
In this study, we consider Wedge as an example and refer to it as the optimized figure \textit{OptWedge} and heuristic parameterized figure vanilla Wedge (\textit{VW}).

The proposed method consists of two parts: modeling of the cognitive cost and solving an optimization problem.
We first define cognitive cost with bias and individual differences.
However, because of the complexity of cognitive processing, it is difficult to formulate the bias and individual differences from a cognitive perspective.
To overcome this, we collected data from an experiment.
We showed many Wedges with different parameters and modeled the obtained data using machine learning techniques.
Subsequently, we minimized the cost to seek the figure, resulting in the low negative impact.
Here, we further investigated that how to handle bias effect the localization accuracy.
This includes two types of OptWedge: \textit{unbiased} OptWedge (\textit{UOW}), which considers bias as undesirable one, and \textit{biased} OptWedge (\textit{BOW}), which consider bias as an available resource for accurate guidance.
We assumed biased OptWedge may be more effective when guiding far POIs and verified whether this hypothesis was valid thorough experiments.
To summarize, our contributions are as follows:

\begin{itemize}
    \item We formulated an optimization problem that ensures both the validity of parameters and scalability of constraints.
    \item We proposed a new kind of cost that considers the cognitive influence to guide off-screen POIs more accurately.
    \item We compared two kinds of optimizations with different parameters and made clear an effect of dealing with the bias.
\end{itemize}


%% file: 2-relatedwork.tex
\section{Related Work}
\subsection{Taxonomy of Off-screen Visualization}
Wedge is an example of \textbf{contextual cue} visualization for off-screen visualization.
Here, we introduce other visualization techniques and clarify the features of the contextual cue technique.
Generally, off-screen visualization methods consider displaying two views that provide information with different concreteness levels: concrete \textit{focused view} and abstract \textit{contextual view}.
Because the two views have a complementary relationship, it is not sufficient to consider one of the views for some tasks.
In a navigation system, users determine their positions via a focused view, while checking the distance to a destination via a contextual view.
Cockburn et al.~\cite{Cockburn09} categorized mechanisms that integrate two views into four schemes, including the contextual cue.
The rest of the schemes are as follows.

\noindent
\textbf{Overview+Detail}~\cite{Javed12}\cite{Kaser11}\cite{Lieberman94}\cite{Stoakley95}\cite{Ware95} spatially separates the two views and displays them side by side.
Typically, the more critical view occupies most of the screen, whereas the other view uses the rest.
Users need to repeatedly look at the two views to assimilate them.

\noindent
\textbf{Zooming}~\cite{Bederson94}\cite{Igarashi00} temporally separates the two views, and smoothly switches them gradually.
The entire display area can be dedicated to one of the views in exchange for a cost to memorize the first view until the assimilation of the second view.

\noindent
\textbf{Focus+Context}~\cite{Brosz13}\cite{Elmqvist08}\cite{Gruenefeld18-2}\cite{Gustafson07}\cite{Ion13}\cite{Jo11}\cite{Sarkar92} distorts the two views to seamlessly blend them spatially, such as the fish-eye lens.
A load associated with assimilation is smaller than the view-separated visualization schemes. 
On the other hand, it is harmful to perceive distance and direction because of the spatial distortion.

Meanwhile, the contextual cue uses proxies that represent information of interest in the contextual view.
In other words, they selectively highlight a particular type of information (e.g., POIs) while eliminating other redundant information, unlike other schemes.
Because this approach saves space for the contextual view, not sacrificing the focused view, the contextual cue is particularly useful for small-sized displays.
Moreover, the contextual cues can be used alongside other schemes~\cite{Miau16}.

\subsection{Digging into Contextual Cue for POIs}
As presented in the previous section, contextual cue techniques provide proxies that selectively highlight the information of interest.
This selection is based on specific search criteria dependent on an application.
To visualize the direction to an object of interest, a simple arrow~\cite{Chittaro04}\cite{Tonnis06} or equivalent~\cite{Wonner13}\cite{Zellweger03} would be suitable as a proxy.
Otherwise, to grasp the situation of interest, the proxy can be something like a mirror that reflects the out-of-view regions~\cite{Fan14}\cite{Lin17}.

Here, we narrow down the search criteria to the POI and compare some proxies for it.
The main task is to encode a distance to the POI into the proxy.
Many proxies express the distance through their relative change.
When one LED becomes brighter, we feel close to a POI corresponding to the LED~\cite{Muller14}.
This relative change also includes an animation~\cite{Gruenefeld18-4}, color~\cite{Gruenefeld18-3}, and vibration~\cite{Stratmann18}.
However, these methods require a long time to form a mental model to intuitively perceive the distance.

On the other hand, Halo~\cite{Baudisch03} and Wedge~\cite{Gustafson08-1} need little time for users to get used to because they utilize an innate cognitive ability referred to as amodal completion instead of forming the mental model.
Wedge was proposed to save the screen space required for displaying the Halo.
Recently, Wedge has become more popular than other proxies.
Because the original Wedge was intended for POIs on a 2D mobile screen, many studies have applied Wedge to guide POIs in 3D virtual~\cite{Gruenefeld18-1}\cite{Gruenefeld17}\cite{Matthias11}\cite{Yu20} or real~\cite{Adcock13}\cite{Petford19} environments.
This popularity may come from the superior effectiveness of Wedge.
In a user study, Yu et al.~\cite{Yu20} found that Wedge was better in distinguishing the distances among multiple POIs.
Petford et al.~\cite{Petford19} also showed that Wedge requires less time for estimation than other proxies.
Although Wedge is not suitable for every task (for example, finding the nearest one from some POIs~\cite{Burigat06}), we believe that it is worth considering the use of Wedge.
Therefore, we selected Wedge in this study and proposed a method to improve its effectiveness; however, our methods are also applicable to other proxies specialized in POI.



%% file: 3-method.tex
\section{Method}
\subsection{Vanilla Wedge}
Wedge has two shape-related parameters: aperture $\theta$ and leg $l$, as shown in Figure~\ref{fig:param}.
Although there is a missing parameter for rotation to prevent overlapping each other, we will consider a simple case where the perpendicular bisector of Wedge is orthogonal to the display boundary.
When the $\mathrm{d_\mathrm{POI}}$ represents the distance from the POI to its nearest point on the screen edge (the origin in Figure~\ref{fig:param}), Gustafson et al.~\cite{Gustafson08-1} heuristically determined parameters based on a constant value $\mathrm{d_{POI}}$ as follows:

\figparam
\formulaoriginal

They also defined ``orbital'' as a region where users possibly estimate.
According to the paper, this term is derived from chemistry, where a molecular orbital is a region in which an electron is found in a molecule, that is, a space with a probability distribution.
It is necessary that the orbital should be small and include the POI inside for achieving more accurate localization.
Gustafson conducted an additional experiment in his thesis~\cite{Gustafson08-2} to evaluate the relationship between the shape of the VW and orbital.
However, its evaluation is too simple to model with high accuracy.
Taking his work as a starting point, we modeled the relationship with high accuracy and applied it to shape optimization.

\subsection{OptWedge}
Solving an optimization problem that minimizes the cognitive cost provides an OptWedge.
We propose two types of OptWedge: unbiased OptWedge (UOW) and biased OptWedge (BOW).
To obtain each OptWedge, we minimize the following Lagrangian that consists of the cognitive cost term $f$ (detailed later), and constraint terms $g_i$.

\formulaopt

Stable optimization requires constraints for the domain of definition ($0<\theta<\pi,0<d<l\cos(\theta/2)$).
Additionally, the developer may add constraints depending on the task.
In the later experiment, we introduced two optional constraints that limit the size of the Wedge so that it is not overlapped with on-screen contents.
Specifically, we consider a bounding box (shaded blue regions in Figure~\ref{fig:optwedge}) of Wedge parameterized with width $w$ and height $h$ and set maximum value as $\mathrm{W}$ and $\mathrm{H}$, respectively.

\subsubsection{Cognitive Cost}
We first describe a cognitive cost that considers the cognitive influence for accurate localization.
This cost measures how desirable the orbital is for achieving localization with high accuracy.
Because the orbital shape is unclear, we approximated the orbital with a normal distribution (red circle in Figure~\ref{fig:param}), which is parameterized using two cognitive factors: bias $b$ and individual differences $\sigma$.
The bias is the distance from the invisible vertex to the mean of the normal distribution.
It is worth noting that the bias takes a positive value if the POI is farther than the invisible vertex and takes a negative value in the reverse situation.
The individual differences appear the standard deviation of the normal distribution.
For simplicity, we consider a normal distribution with independent dimensions, that is, the individual differences have two components $\sigma = (\sigma_x, \sigma_y)$ for a 2D Wedge.
If we denote $d$ as the distance from the origin to the invisible vertex, we can write a normal distribution $P$ as a function of the parameters $(\theta, l, d)$.

\formulap

Subsequently, we consider an ideal normal distribution $\mathrm{Q}$.
When the mean of the normal distribution is on the POI, and the standard deviation is zero, it can be inferred that the normal distribution is ideal because all humans can determine the POI with no mistakes; this normal distribution is expressed below.
The standard deviation $\mathrm{\epsilon_x}$ and $\mathrm{\epsilon_y}$ takes small enough constant values.

\formulaq

The cognitive cost represents how far the two distributions $P$ and $\mathrm{Q}$ are.
Although there are many possible ways to numerically express this gap, it needs to be suitable for an application.
For example, we can introduce a new hyperparameter that gives a larger penalty to the distance error than the direction error. 
For simplicity, we utilized the Kullback-Leibler divergence~\cite{Kullback51} to measure the gap because it does not require extra hyperparameters.
This divergence $f$ takes a zero value when the distributions match and a positive value otherwise.

\formulakld

\subsubsection{UOW and BOW}
The difference between the UOW and BOW is the optimization of distance $d$.
Figure~\ref{fig:opt} provides an intuitive understanding.
Main difference is whether an invisible vertex is on a POI or not.
In other words, distance $d$ is equal to $d_\mathrm{POI}$ for UOW, and it is not for BOW.
BOW seems like a trick because the user believes the POI is on the invisible vertex, although it is not.
However, we deduce that this trick works because we could estimate a potential bias in advance using the model $P(\theta, l, d)$.
To summarize, the UOW and BOW differ in the way they handle bias.
The UOW considers bias as an undesirable product of cognitive processing and makes the potential bias close to zero while optimizing, whereas the BOW does not.
We assumed that the BOW worked better than the UOW when the POI was far from the screen.
As the POI gets farther away, it becomes more challenging to find parameters with a small bias.
To validate this hypothesis, we compare the UOW and BOW performances at different distances to the POI in a later experiment.

\figopt


%% file: 4-evaluation.tex
\section{Experiments}
\subsection{Experiment 1: Modeling}
\subsubsection{Setting}
The goal of the first experiment is to model the cognitive cost required for the second experiment.
To collect a lot of data necessary for the regression, we presented many Wedges with different parameters and asked participants to estimate the invisible vertex.
After a pilot test, we determined the range of each parameter as follows: $\theta\ [^\circ]\in\left\{10,30,\ldots,130, 150\right\}, l\ [\mathrm{m}]\in\{2,3,\ldots,11,12\}, d\ [\mathrm{m}]\in\{1,2,\ldots,10,11\}$ .
It should be noted that these values are based on a distance of $10~\mathrm{m}$ from the participant to the screen.
There were a total of 968 possible combinations, but only 375 were valid combinations that satisfy the domain of definition.

The experiment was conducted in virtual reality (VR) environment for two reasons.
The first reason is to validate the effectiveness of our method for a small-sized screen, such as head-mounted displays (HMD).
The second reason is that the participants can easily input their guessed positions using a controller.
They move their arm so that a ray emitted from the controller hits the point estimated as POI.
This interaction allows us to collect more data in less time.
However, because of the difficulty in grasping the distance in the VR environment, we additionally showed a grid on the screen.

This and the next experiments were approved by the Mitsubishi Electric and conducted according to the principles of Declaration of Helsinki.
Following a brief explanation of the task, participants sat in a chair and wore HTC Vive Pro\footnote{https://www.vive.com/jp/product/vive-pro-full-kit/}.
We then allowed the participants to practice the operation for approximately one min.
Subsequently, we displayed many Wedges in random order and rotation angles while having a five-minute break for every 100 answers.
This procedure was conducted in the same way for 15 male and 5 female participants, with their age groups ranging from 20s to 50s.
All participants had an average visual ability.

\figraw
\figoffline

\subsubsection{Results}
We calculated bias $b$ and individual differences $\sigma_x$ and $\sigma_y$ from raw data after removing some outliers (blue crosses in Figure~\ref{fig:trend}) using Hotelling's $T^2$ method\footnote{We set the significance level at 5\% for anomaly detection.}~\cite{Hotelling31}.
Figure~\ref{fig:raw} shows the results.
We compared some examples shown in Figure~\ref{fig:trend} and found the following T1 and T2 trends.

\noindent
\textbf{T1} When a POI is far from a screen, bias takes a negative value, that is, humans tend to underestimate the distance to the POI (see Figure~\ref{fig:trend} (a)).

\noindent
\textbf{T2} Concerning the aperture $\theta$, there is an error trade-off between individual differences $\sigma_x$ and $\sigma_y$ (see Figure~\ref{fig:trend} (b)).

Then, we divided dataset into training set ($80\%$) and test set ($20\%$) and modeled with 5-folds cross validation.
For each cognitive factor, we performed polynomial regression (PR) and Gaussian process regression (GR).
For the PR models, we have compared models with different orders while previous studies~\cite{Gustafson08-2} have tested linear regression models.
We added the $L_2$ regularization term into each model to avoid over-fitting.
For the GR models, we adopted a combination of the Mattern 2/5 kernel and a linear kernel and determined the kernel parameter values based on the validation dataset.

We first evaluated the coefficient of determination $R^2$ according to the previous analysis.
We did not evaluate the GR model because we employed the adjusted $R^2$, where the number of explanatory variables is required to compare performance among different orders.
Figure~\ref{tab:r2} represents the comparison in the test dataset.
We can see that our PR models outperformed the preceding models, and found the best order suitable for modeling each cognitive factor.
We then calculated the mean squared error (MSE) to evaluate the GR models.
Figure~\ref{tab:mse} presents the results of comparing our PR models and GR models in the test dataset.
For each PR model, we selected the best order in Figure~\ref{tab:r2}; we used a quadratic model for $b$ and $\sigma_y$ and a linear model for $\sigma_x$.
As indicated by Figure~\ref{tab:mse}, the GR models are suitable for modeling all cognitive factors with high accuracy.
Therefore, we employed the GR models for calculating the cognitive cost when optimizing it in the next experiment.

\tabreg

\subsection{Experiment 2: Optimization}
\subsubsection{Setting}
The goal of the second experiment is to compare OptWedge with vanilla Wedge to determine its effectiveness.
The base setting is common to the first experiment.
We asked participants to perform the same task using the same procedure.
However, we increased the number of times we showed Wedge with the same parameter from once to twice to remove the noise included in the obtained data.
We also increased the number of participants from 20 to 22 for the same reason.

In this experiment, We presented three types of Wedge, the VW, UOW, and BOW.
We set the distance $d_\mathrm{POI}\mathrm{[m]}\in\{1, 2, \ldots,10, 11\}$ and generated each Wedge in advance.
When optimizing the UOW and BOW parameters, we iteratively performed a gradient descent starting from the VW parameters as the initial point.
We hoped this initialization would give desirable parameters with a lower cognitive cost than VW.
Each constant value is set as follows: To keep the Wedge within the field of view of the HMD, we set the maximum size of the drawable area as $\mathrm{W} = \mathrm{H} = 14~\mathrm{m}$.
For the ideal normal distribution $\mathrm{Q}$, we set $\mathrm{\epsilon_x^2} = \mathrm{\epsilon_y^2} = 0.1~\mathrm{m^2}$. 

\figresults

\subsubsection{Results}
We conducted quantitative and qualitative assessments on the obtained data.
For the quantitative assessments, we evaluated the cognitive cost to validate the accuracy of the models created in the previous experiment.
Figure~\ref{fig:comparison} compares the costs of each Wedge with respect to the different distances $d_\mathrm{POI}$. 
The solid lines represent the actual cost calculated from the obtained data, whereas the dotted lines show the model predictions.
The results show that the closer the distance $d_\mathrm{POI}$, the lower the cost of the UOW and BOW tends to compared to that of VW.
However, as the distance increases on the border of $d_\mathrm{POI} = 7~\mathrm{m}$, it appears that the model is no longer accurate and loses the advantage of OptWedge.

To clarify the effectiveness of OptWedge at a distance of $d_\mathrm{POI} < 7~\mathrm{m}$, we also evaluated the root mean square error (RMSE) from the POI to each plotted point.
Figure~\ref{fig:test} shows the results of Wilcoxon signed-rank test considering Bonferroni correction.
We identified significant differences between VW and UOW and VW and BOW at $d_\mathrm{POI} = 1~\mathrm{m}$ and $2~\mathrm{m}$.
In the other case, although we found no significant differences, the mean and variance of the UOW tended to be lower than that of VW.

As a qualitative assessment, Figure~\ref{fig:optwedge} visualizes the cognitive cost in the parameter space and Wedge corresponding to points in the parameter space.
From this visualization, we can see that the UOW has a global minimum solution within the region bounded by the black lines corresponding to the constraints.
It should be noted that because the cost visualization is a clipped parameter space for $d=d_\mathrm{POI}$, there is no blue point for the BOW in the parameter space.
The results show that the optimization has the following effects: E1 and E2.

\noindent
\textbf{E1} The apertures $\theta$ in the UOW and BOW are larger than for VW (except for $d_\mathrm{POI}=11~\mathrm{m}$).

\noindent
\textbf{E2} The invisible vertex of the BOW is farther than the POI at $d_\mathrm{POI}=9~\mathrm{m}$ and $11~\mathrm{m}$.

\figoptwedge

%% file: 5-discussion.tex
\section{Discussion}

\subsection{Effectiveness}
We categorize the results into three groups: near $(d_\mathrm{POI} \leq 2~\mathrm{m})$, medium $(2~\mathrm{m} < d_\mathrm{POI} \leq 7~\mathrm{m})$, and far $(7~\mathrm{m} < d_\mathrm{POI} \leq 11~\mathrm{m})$.
More generally, if we consider the viewing angle, these values correspond to less than or equal to approximately $11^\circ$, $35^\circ$, and $52^\circ$, respectively.

\noindent
\textbf{Near}
We think the effect E1 explains why UOW and BOW lead to more accurate localization than VW at short distances.
Because E1 is consistent with the hypothesis from the previous study~\cite{Gustafson08-1} that ``larger apertures would have led to smaller orbitals,'' we can assume that this effect improves the human estimation.
Although we could not confirm this effect at $\mathrm{d_\mathrm{POI}}=11~\mathrm{m}$, we consider it to be an exception because the model is not accurate at long distances.
Figure~\ref{fig:comparison} shows that the model predictions deviate from the experimental cost for the BOW.

\noindent
\textbf{Medium}
No significant difference was obtained in the medium distance because the accuracy of VW was sufficiently high.
Observing the initial point (VW) and optimal point (UOW) in the parameter space in Figure~\ref{fig:optwedge}, those points are getting closer to each other as the $\mathrm{d_\mathrm{POI}}$ increases. 
This observation means that the previous method~\cite{Gustafson08-1} has a certain validity despite being heuristic.
However, because VW has difficulty dealing with the constraints explicitly, we recommend using the UOW, which works most stably.

\noindent
\textbf{Far}
We cannot agree with our hypothesis that the BOW is superior to the UOW at a long distance because our model becomes inaccurate as the distance to the POI increases.
One possible cause of this fact is a lack of data.
Figure~\ref{fig:raw} shows that the number of parameter combinations satisfying the domain of definition decreases as the distance $d$ increases.
Because we set the parameter values as equally spaced, the number of data points around a long distance is insufficient for improving the generalizing capability.
To avoid this problem, in the first experiment, it is desirable to adaptively change Wedge parameters based on the user response, such as active learning, instead of equal spacing.
Alternatively, we should have presented the Wedge with the same parameters multiple times to model not only the variance between the individual but also the variance within individuals.
However, we think the effect E2 is reasonable because it comes from learning the trend T1.
Learning the T1 indicates that an optimizer input larger $d$ than $d_\mathrm{POI}$ to the model.
We inferred that the BOW tried to consider the cognitive influence but failed because of the generalizing capability in the model.

\subsection{Limitations and Future work}
\textbf{Robustness}
From Figure \ref{fig:optwedge}, we can see that our model has lack of robustness at long distance.
For (d) and (e) in the figure, BOW is very close to the edge, potentially making it difficult to judge distance.
This is because modeling the trend T1 makes the optimization tried to adopt bigger $d$ value in spite of the lack of data for $d > 11~\mathrm{m}$.
To improve robustness, we will introduce continuity constraints of optimized parameters as $d$ changes.

\noindent
\textbf{Generalizability} 
The above classification and our models may be specialized in VR environments.
We observed that the user usually moves the line of sight and the head when estimating a more distant POI in experiments.
When users move their heads, they can not see the displayed portion of Wedge because of the narrow viewing angle of the HMD, leading to interference with amodal completion.
The viewing distance from the participant to the screen was also fixed, which may affect the generalizability of the results as different distances can lead to different sizes of the visualization.
Therefore, we will conduct the same experiments in various environments and comparatively evaluate the results in the future.

%% file: 6-conclusion.tex
\section{Conclusion}
This study introduced a cognitive cost that considers bias and individual differences to improve localization accuracy for guiding off-screen POIs.
We proposed a method to optimize a figure referred to as Wedge.
We also designed two kinds of optimized Wedge (OptWedge) with different handling of bias:
the unbiased OptWedge (UOW) tried to approach bias zero, whereas the biased OptWedge (BOW) does not.
We conducted two experiments.
The first experiment showed that our model is more accurate than the existing model.
The second experiment revealed that OptWedge is valid for relatively close POI, and the validity of vanilla Wedge (VW).
However, it appears that our model may be over-fitting and specialized in a VR environment.
We will tackle these limitations and apply our idea to other visualization techniques.